\documentclass[11pt,draftcls,onecolumn]{IEEEtran}
\usepackage{epsfig}
\usepackage{dsfont}
\usepackage{amsmath}
\usepackage{amstext}
\usepackage{amsfonts}
\usepackage{amssymb}
\usepackage{eucal}
\usepackage{graphicx}
\usepackage{verbatim}
\usepackage{stmaryrd}
\usepackage{bm}

\usepackage{tikz}
\usepackage{pgfplots}
\usepackage{algorithmic}

\usetikzlibrary{arrows,petri,topaths}
\newcommand{\trans}{{\sf T}}

\newcommand{\x}{{\bf x}}

\newcommand{\EE}{{\rm E}}

\newcounter{cdefinition}
\newtheorem{definition}[cdefinition]{Definition}

\begin{document}
\bibliographystyle{IEEEtran}

\title{Electrical Vehicles in the Smart Grid: \\ A Mean Field Game Analysis} 
\author{	Romain~Couillet,
	    	Samir M.~Perlaza,    %~\IEEEmembership{Student Member,~IEEE,}
        		Hamidou Tembine,%~\IEEEmembership{Member,~IEEE,}
        		~and~M\'erouane Debbah   %~\IEEEmembership{Senior~Member,~IEEE}% <-this % stops a space
\thanks{R. Couillet is with SUPELEC. $3$ rue Joliot-Curie, $91192$, Gif-sur-Yvette, cedex. France. (romain.couillet@.supelec.fr) }
\thanks{S. M. Perlaza is with Alcatel-Lucent Chair in Flexible Radio at SUPELEC. $3$ rue Joliot-Curie, $91192$, Gif-sur-Yvette, cedex. France. (samir.medina-perlaza@supelec.fr)}%
\thanks{H.~Tembine is with  the Telecom Dept. at SUPELEC. $3$ rue Joliot-Curie,  $91192$, Gif-sur-Yvette, cedex. France. (hamidou.tembine@supelec.fr)}%
\thanks{M. Debbah is with Alcatel-Lucent Chair in Flexible Radio at SUPELEC. $3$ rue Joliot-Curie, $91192$, Gif-sur-Yvette, cedex. France. (merouane.debbah@supelec.fr)}%
%\thanks{The material in this paper has been presented in part in the IEEE Intl. Workshop on Signal Processing Advances for Wireless Communications in 2010 (SPAWC 2010) \cite{Perlaza-Spawc2010}}.
}

\maketitle
\begin{abstract}
In this article, we investigate the competitive interaction between electrical vehicles or hybrid oil-electricity vehicles in a Cournot market consisting of electricity transactions to or from an underlying electricity distribution network. We provide a mean field game formulation for this competition, and introduce the set of fundamental differential equations ruling the behavior of the vehicles at the feedback Nash equilibrium, referred here to as the mean field equilibrium. This framework allows for a consistent analysis of the evolution of the price of electricity as well as of the instantaneous electricity demand in the power grid. Simulations precisely quantify those parameters and suggest that significant reduction of the daily electricity peak demand can be achieved by appropriate electricity pricing.
\end{abstract}

\section{Introduction}
\label{sec:intro}
Electrical vehicles (EV) and plug-in hybrid electrical vehicles (PHEV) have been recognized as natural components of future electricity distribution networks, known as smart grids \cite{Simpson-2006, Bradley-2009, KIN07}. As opposed to classical vehicles, EV and PHEV are equipped with batteries which can be charged or discharged by using a simple plug-in connector compatible with the local electricity distribution grid. Thus, EV and PHEV can be conceived as both energy consuming devices and mobile energy sources  \cite{Pan-2010,  Shao-2009, Sojoudi-2011, Turton-2008}. In the former case, EV and PHEV can be seen as devices straining the energy demand of energy suppliers and, thus, adding a new constraint to reliably distribute the electricity. In the latter case, EV and PHEV can be used to store or even to transport the energy from one geographical area to another and then to increase the reliability of the energy supply in certain zones or time intervals. 

In this framework, it is therefore an important economical and social challenge to enforce charge and discharge policies to EV and PHEV in an optimal manner. Here, optimality must be interpreted in the sense of individual revenue obtained by the EV and PHEV owners when participating in the energy trades and also in terms of reliability of the energy supply process to the fixed consumers.
In this paper, we consider that a way to improve reliability is to allow EV and PHEV to buy and sell energy to or from the smart grid, as in a classical Cournot competition \cite{Cournot-1929}. Clearly, the price at which the energy is sold and bought depends on the existing demand in the grid and also on the demand and offer resulting from all the vehicles connected to the network. This competitive interaction resulting from the energy trade, given a global price, can be analyzed using tools from dynamic game theory \cite{Basar-Book}.
This is studied for instance in \cite{Saad-2011}, where a noncooperative game is played among a number of PHEV groups aiming to sell part of their stored energy to the smart grid; an algorithm based on best response dynamics is then proposed to allow PHEV groups to reach a Nash equilibrium.

Nonetheless, in practical scenarios, the number of vehicles might be drastically large so that finite dimensional game theory analysis might not necessarily bring enough insight about the global behavior of the market. 
To overcome this problem, in this paper, we study the energy trade when the number of vehicles tends to infinity and all vehicles are considered alike, following the paradigm of \cite{Aumann-1964,Carmona-2004}. More precisely, we shall model this interaction as a mean field game \cite{LAS06,LAS06b}.
In contrast to finite games, where each player follows the evolution of the state of the game and the actions taken by all other players in order to maximize a given individual benefit, in the mean field game formulation, players do not react to actions from individual players but rather to the aggregate behavior of all players. 
The notion of (Nash) equilibrium in the context of mean field games is known as mean field equilibrium (MFE). When focusing only on the class of regular functions of time and battery levels, a necessary condition for the MFE is to be the solution of a coupled system of partial differential equations (PDE) which includes a (backward) Hamilton-Jacobi-Bellman (HJB) equation and a (forward) Fokker-Planck-Kolmogorov (FPK) equation. %The HJB equation determines the optimal action for each player for a given density measure on the set of possible states of the players. The FPK determines the density of the invariant measure of the (ergodic) game state  for a given action or  control. 

The closest contribution to our specific problem setting is \cite{GUE10,GUE11}. Therein, a mean field game approach to the study of oil production is developed. In \cite{GUE10}, the selfish players are oil producers and the mean field variable is the oil selling price. In this article, we develop a similar framework as in \cite{GUE10} but on a finite time horizon, applied to both EV and PHEV, with vehicle owners as the selfish players and electricity price as the mean field variable.  
%
%In general, it is difficult to prove that the mean field solutions are well-defined limits of the finite game, see e.g. the discussions in \cite{GUE11}. Similarly, it is in general difficult to prove the existence and uniqueness of solutions of one equilibrium and, if so, to derive a numerical method that is provably able to converge to an equilibrium, see e.g. \cite{ACH10}. We will not try to prove any of these aspects here, our main target being rather to informally explore the potentials of the mean field game setting to the economical problem of optimal policies for EV and PHEV penetrations in the smart grid. Similar to \cite{GUE10}, only the numerical results will convey a justification of the correct behavior of our method.

The reminder of this article unfolds as follows. In Section \ref{SecPurelyEV}, we describe the problem formulation in the case where only electrical vehicles interact with the energy market. Therein, the problem is formulated as a continuous time differential game with finite time horizon. This formulation is then written under the form of a mean field game and the differential equations describing the MFE are presented. In Section \ref{SecPHEV}, the same analysis presented for EV is carried out for the case of PHEV. In Section \ref{SecSimulations}, we provide numerical simulations and derive conclusions for both scenarios. Finally, in Section \ref{SecConclusion}, we conclude this work.

\section{Electrical Vehicles}\label{SecPurelyEV}

\subsection{System Model}
Consider a finite set $\mathcal{K} = \{1,\ldots,K\}$ of EVs participating to energy trading with an underlying electricity distribution network. The consumption rate of vehicle $k \in \mathcal{K}$ at time $t\in [0,T]$ is denoted by $g_{t}^{(k)}$. This consumption rate is measured in units of electricity per time. We assume that $g_{t}^{(k)}$ is deterministic and known by EV $k$. %We also denote $g^{(k)} = \{ g_{t}^{(k)}, 0 \leq t \leq T \}$.
The amount of energy stored in the battery of vehicle $k$ at time $t$ is denoted by $x_{t}^{(k)} \in [0,1]$, quantified in energy units. Here, $x_{t}^{(k)}=0$ for an empty battery and $x_{t}^{(k)}=1$ for a fully charged battery.
We denote by $\alpha_t^{(k)}$ the energy provisioning rate of vehicle $k$ at time $t$, that is, the rate at which vehicle $k$ buys or sells its energy.  
We relate the variable $x_{t}^{(k)}$ to $g_{t}^{(k)}$ and $\alpha_{t}^{(k)}$ by the following differential equation
\begin{equation}
	\label{EqStateODE1}
\frac{{\mathrm d}}{{\mathrm d} t}x_t^{(k)} = \alpha_t^{(k)} - g_t^{(k)},
\end{equation}
where $\alpha_t^{(k)}$ and $g_t^{(k)}$ are chosen such that the trajectory $x_t^{(k)}$ is unique for a given initial $x_0^{(k)}$ and that, for all $t$, $0 \leq x_t^{(k)} \leq 1$. Such $\alpha_t^{(k)}$ is called an {\it admissible} provision rate.

In the following, we denote $\bm{x}_t = ( x_t^{(1)}, \ldots, x_t^{(K)})$ and $\bm{\alpha}_t = ( \alpha_t^{(1)}, \ldots, \alpha_t^{(K)})$ the battery level profile and provisioning rate profile at time $t$, respectively. Consider now a predefined period $[0,T]$. We denote $x^{(k)} = \{ x_t^{(k)}, 0\leq t \leq T\}$ and $\alpha^{(k)} = \{\alpha_t^{(k)},0\leq t \leq T\}$ the trajectories of the battery level and provisioning rates for EV $k$, respectively. We also denote $\bm{x}=\{\bm{x}_t,0\leq t\leq T\}$ and $\bm{\alpha}= \{\bm{\alpha}_t,0\leq t \leq T\}$ the trajectories of the battery level and provisioning rate profiles. We finally denote $\mathcal A^K$ the set of all admissible provision rates $\bm{\alpha}$.

The price at which vehicles either sell or buy electricity at time $t$ is determined by the function $p_t: \mathds{R}^{K} \rightarrow \mathds{R}$, $\bm{\alpha}_t\mapsto p_t(\bm{\alpha}_t)$. The time dependency of the price $p_t$ models a realistic dynamic pricing policy accounting for the energy demand for other services than EV battery loading. This function can be tuned to create incentives for EV to sell or buy energy at specific time periods. In addition to electricity price, other factors influence the energy trades of EV owners. We model the latter, for player $k$, by the following set of functions.
The function $h_{t}^{(k)}: \mathds{R}\rightarrow \mathds{R}$, $\alpha\mapsto h_{t}^{(k)}( \alpha)$ models the (psychological) cost for player $k$ to buy or sell electricity at rate $\alpha$ at time $t$. Indeed, EV owners are more likely to trade energy at some convenient time intervals, e.g. during nighttime when the EV is parked at home. The function $f_{t}^{(k)}:[0,1]\to \mathds{R}$, $x\mapsto f_{t}^{(k)}(x)$ models the cost for vehicle $k$ to possess only a fraction $x$ of energy reserves at time $t$. For instance, during periods of high energy consumption, the interest of EV owners is to have maximally loaded batteries.
Finally, $\kappa^{(k)}: [0, 1] \rightarrow \mathds{R}$, $x\mapsto \kappa^{(k)}(x)$ models the cost for EV $k$ to end the trade period $[0,T]$ with a fraction $x$ of battery load. This function guarantees that EV owners do not sell all their battery content at the end of the trade. A comprehensive discussion on the choices of these functions is considered in Section \ref{SecSimulations}.
 
The goal of EV $k$ is to determine the consumption rates $\alpha^{(k)}$ that minimize its total cost $J_k: \mathcal{A}^K  \rightarrow \mathcal{R}$, $(\alpha^{(1)}, \ldots, \alpha^{(K)}) \rightarrow J_k (\alpha^{(k)},\bm{\alpha}^{(-k)})$, over a time window $[0,T]$ given the consumption rates $\bm{\alpha}^{(-k)}$ chosen by all the other EVs. That is, 
\begin{eqnarray}
\label{EqCostFunctionEV}
J_k\left(\alpha^{(k)},\bm{\alpha}^{(-k)}\right)	& = & \displaystyle\int_0^T \left( \alpha_t^{(k)} p_t(\bm{\alpha}_t) + h_t^{(k)}(\alpha_t^{(k)}) + f_t^{(k)}(x_t^{(k)})\right) {\mathrm d} t +  \kappa^{(k)} (x_T^{(k)})
\end{eqnarray}
for a given initial state $\bm{x}_0$.
%where ${\alpha}^{(k)}_t$ is taken here to be a function of the {\it own state} $x_t^{(k)}$ and of the global price $p_t(\bm{\alpha}_{t})$ at which the energy is being traded. 
Note importantly that the instantaneous global price $p_t(\bm{\alpha}_{t})$ is a function of  the instantaneous provisioning rate profile $\bm{\alpha}_{t}$, which in return depends both on the instantaneous energy reserve profile $\bm{x}_{t}$ and on the initial energy reserve profile $\bm{x}_{0}$. 

In the following, we formulate a differential game which models the interactions between the active EVs in the system.
%%%%%%%%%%%%%%%%%%%%
%%%%%%%%%%%%%%%%%%%%

\subsection{Classical Game Formulation}\label{SecDiffGame1} 

We model the energy trades resulting from the interactions among the electrical vehicles and the smart grid by a $K$-player continuous-time differential {\it game} of pre-specified fixed duration $T>0$. 
Let $\mathcal{K}$, the set of EV, be the set of {\it players}. %Here, the set of control functions or simply the controls of player $k$ is the set $\mathcal{A}_k$ described above. 
The state of the game, at time $t$, is determined by the energy reserve profile $\bm{x}_t = ( x_t^{(1)}, \ldots, x_t^{(K)})$, whose trajectory $\bm{x}$ is determined by the initial state $\bm{x}_{0}$ and, through the players' control, by the state evolution equation \eqref{EqStateODE1}.  
The cost function of player $k$ is defined by \eqref{EqCostFunctionEV}. The objective of player $k$ is to determine a control trajectory $\alpha^{(k)}$ that minimizes its cost. At instant $t$, the instantaneous control $\alpha_{t}^{(k)}$ is determined based on the information available to player $k$, which we denote by the information set $\eta_{t}^{(k)}$. We will consider here that the information set corresponds to the singleton $\eta_{t}^{(k)} = \lbrace x_t^{(k)} \rbrace$. That is, players are assumed memoryless as they do not remember the previous individual states nor their previous instantaneous controls.  

 In the following, we describe the strategy of player $k$, that is, the mapping from the individual information set to the space of individual controls. Let us denote the strategy of player $k$ by the mapping $\gamma_{t}^{(k)}: \eta_t^{(k)} \rightarrow \mathcal{A}^K_{t,k}$, $x\mapsto \gamma_{t}^{(k)}(x)$ with $\mathcal{A}^K_{t,k}$ the set of admissible controls for player $k$ at time $t$. Given the nature of the information sets, this strategy can be referred to as a {\it non-anticipative (own-state) feedback strategy}. The action of player $k$ is therefore described as
\begin{equation}\label{EqAlphaDef}
\alpha_t^{(k)} =  \gamma_{t}^{(k)}(x_t^{(k)}).
\end{equation}
In the following, we will mostly use the notation $\alpha_t^{(k)}$, implicitly assuming the existence of a mapping $\gamma_{t}^{(k)}$, and we denote $\bar{\mathcal A}_t^K=\bar{\mathcal A}^K_{t,1}\times \ldots\times \bar{\mathcal A}^K_{t,K}$, $\bar{\mathcal A}^K_{t,k}\subset \mathcal A^K_{t,k}$, the class of feedback strategies at time $t$. The notation $\bar{\mathcal A}^K\subset \mathcal A^K$ will be used for the class of all $K$-vector feedback controls $\{\bar{\mathcal A}_t^K,0\leq t\leq T\}$. 
We recall that the interdependence between players in this game appears through the electricity price: the individual control $\alpha_{t}^{(k)}$ depends on the global price $p_t(\bm{\alpha}_t)$, which depends itself on all the other players' individual controls $\bm{\alpha}_{t}^{(-k)}$.

The formulation of the game is completed by further imposing that both the deterministic function $g_t$ and the corresponding strategies $\gamma^{(1)},\ldots,\gamma^{(K)}$, with $\gamma^{(k)} = \lbrace \gamma^{(k)}_t: 0 \leq t \leq T \rbrace $, are such that the trajectory defined by the initial value $\bm{x}_0$ and the differential equation \eqref{EqStateODE1} is well defined and unique. 

Following the above game formulation, we consider as equilibrium notion the own-state feedback Nash equilibrium, which we define as follows.
\begin{definition}\label{DefNE} The control profile $\bm{\alpha}^\star = \left( \alpha^{\star(1)}, \ldots, \alpha^{\star(K)}\right) \in \bar{\mathcal{A}}^K$ is an {\it own-state feedback Nash equilibrium} (NE) if, for all $k \in \mathcal{K}$ and for all admissible control $\bm{\alpha} = \left( \alpha^{(1)}, \ldots, \alpha^{(K)}\right) \in \bar{\mathcal{A}}^K$, it holds that
\begin{equation} 
	J_k\left( \alpha^{\star(k)}, \bm{\alpha}^{\star(-k)}\right)  \leq J_k\left( \alpha^{(k)}, \bm{\alpha}^{\star(-k)}\right),
\end{equation}
with $\alpha_t^{\star(k)}=\gamma_{t}^{(k)}(x_t^{\star(k)})$, $\alpha_t^{(k)}=\gamma_{t}^{(k)}(x_t^{(k)})$, $k\in \mathcal{K}$, and $\x_t^\star,\x_t$ satisfying the state evolution \eqref{EqStateODE1}, for a common initial state $\bm{x}_0$.
%for information sets $\eta^{(k)} = \lbrace \eta_t^{(k)}: 0 \leq t \leq T \rbrace$, $\eta_t^{(k)}=\{x_t^{(k)} \}$, $k\in \mathcal{K}$.
\end{definition}

Our interest in the NE lies in the fact that, at a state of NE, all the EV use a control policy, from which they have no reason to depart.
Nonetheless, analyzing the NE of such a game, where $K$ is greater than one
%and in which all players have individual needs 
is a difficult problem.
% which, if it exists
%solvable
In fact, even if a NE exists, it would lead to solutions that are inherently difficult to exploit. In particular, it is clear that, under this formulation, any change in the battery level of a given player impacts all other players which must react as a consequence. We aim at reducing this complexity by adopting some additional, but reasonable, conditions.

\subsection{Mean Field Game Formulation}\label{SecMFGformulationEV}
In this section, we simplify the equilibrium analysis of the differential game presented in the previous section by considering the following hypotheses: (i) the set of players is sufficiently large to be considered infinite, and (ii) players are indistinguishable, in the sense that a different player labelling leads to the same joint state distribution.
The first assumption is tenable here as we analyze a large population of EVs. The second assumption reflects the fact that all players have, to some extent, similar batteries and similar individual objectives (but obviously different battery states). %Note that a practical scenario where both conditions are met is the recent French system of electrical vehicle rental, known as \emph{Auto'Lib} \cite{AutoLib-WebRef}.

 %
%\red{The game described in the previous section can be remarkably simplified by considering a large population of identical players. We assume that the number of players $K$ grows to infinity in a fluid manner, in such a way that the set of players will now be considered continuous and that the individual cost functions become identical. This implies that the set of controls is now the same for all players. 
%We will denote $\mathcal{A}$ this set of control trajectories $\alpha:t\mapsto \alpha_t$ for any player, and we still restrict $\mathcal{A}$ to own-state feedback controls, i.e. $\alpha_t$ is adapted to the state $x_t$ of the player under consideration. In this context, the impact of a particular control adopted by a single individual represents a negligible cost variation on all the other players' costs. Here, players are sensible to the distribution of all other players' states.} 

From the assumptions of player indistinguishability and under the large $K$ limit setting \cite{kotelenez2010macroscopic}, we can drop the player indexes in the previous notations and model the (battery) states of the players at time $t$ by a random variable $x_t$ with distribution $m(t,x)$. As such, $m(t,x)$ is the limiting distribution of the empirical distribution $m^K(t,x)$ defined as
\begin{equation*}
	m^K(t,x) = \frac{1}{K}\sum_{k=1}^K \delta_{x_t^{(k)}=x}.
\end{equation*}

Now, in order to avoid the unrealistic assumption that all vehicles consume energy at the same rate at any time instant, we model the EV consumption rate by the stochastic process $g_t{\mathrm d}t+g_t\sigma_t {\mathrm d}W_t$, with $W_t$ a Brownian motion. The state evolution of $x_t$ is therefore described by the following stochastic differential equation (SDE)
\begin{equation}\label{EqStateSDE}
{\mathrm d} x_t = \alpha_t {\mathrm d} t - g_t \left({\mathrm d}t + \sigma_t {\mathrm d} {W}_t \right) + {\mathrm d} {N}_t,
\end{equation}
with $x_0\in[0,1]$ (now seen as a random variable) having distribution $m_0=m(0,\cdot)$. The term ${\mathrm d} N_t$ is a reflective variable to ensure that $x_t$ remains in $[0,1]$. Similar to above, we will assume that all conditions are met for such a trajectory $x_t$ to be well-defined.
Now, under the assumption of player indistinguishability, the analysis of the game reduces to the study of the trajectory of the individual state and individual control of a single player game (or equivalently, of a stochastic control problem), with cost function $J: \mathcal{A} \rightarrow \mathds{R}$, $\alpha \mapsto J(\alpha)$, with $\mathcal{A}$ the set of all controls $\{\alpha_t,0\leq t\leq T\}$ admissible for the state dynamics \eqref{EqStateSDE}, defined as
\begin{eqnarray}\label{EqCostFunctionEVinfty}
J \left( {\alpha} \right)	& = &  \EE \displaystyle\int_0^T \left( \alpha_t p_t(m_t) + h_t(\alpha_t) + f_t(x_t)\right) {\mathrm d} t +  \kappa (x_T),
\end{eqnarray}
for a given initial $(x_0,m_0)$, where $m_t=m(t,\cdot)$ is the distribution of the players among all individual states, and $x_t$ satisfies the dynamics \eqref{EqStateSDE}. The control $\alpha_t$ is a feedback control that can be seen as the image $\alpha_t=\gamma_t(x_t)$ of the (own-state) feedback strategy $\gamma_t:\eta_t\to \mathcal A_t,x\mapsto \gamma_t(x)$ on the information set $\eta_t=\{x_t\}$. The set of such controls is denoted $\bar{\mathcal A}_t$, and the set of control profiles $\{\alpha_t,0\leq t\leq T\}$ is denoted $\bar{\mathcal A}\subset \mathcal A$.
%The initial state conditions for the considered player are $x_{0} \in [0,1]$ and $m_0\triangleq m(0,\cdot)$, the initial density function.

In this context, the energy trading price writes $p_t: \mathcal{M}_t \rightarrow \mathds{R}_+$, $m_t\mapsto p_t(m_t)$, with $\mathcal M_t$ the class of distributions $m_t$. The price can now be seen as a function of the total instantaneous EV demand $\int_0^1 \alpha_t m_t(\mathrm x)$. However, for computational ease, we will instead consider that prices are fixed not by the total EV consumption $\int_{0}^{1} \alpha_tm_t({\mathrm d} x)$ but by the {\it expected} consumption $g_t + \frac{{\mathrm d}}{{\mathrm d} t}(\int_0^1  x m_t(\mathrm dx))$, where both quantities only differ by an additional Brownian motion term when $\sigma_t>0$. In practice, this suggests that the energy regulators which set the instantaneous prices do not have the information on the instantaneous demand at time $t$ but know the distribution $m_t$ at time $t$ (as we will see, this information is accessible in anticipation at time $t=0$). 
We therefore define $p_t$ as
\begin{equation*}
	p_t( m_t ) = D(t,\cdot)^{-1} \left( g_t + \frac{{\mathrm d}}{{\mathrm d} t} \int_0^1 x m_t(\mathrm d x) \right),
\end{equation*}
where $D(t,p)$ is the total energy demand function (including both EV and external trades) at time $t$ for a given price $p$, and the inverse is with respect to composition.
Under the above assumptions, the continuous time differential game discussed in Section \ref{SecDiffGame1} becomes a mean field game as introduced in \cite{LAS06,LAS06b}.
%This mean field formulation suggests that energy is depleted or filled constantly in the batteries of EV owners. This is justified as we now remind that consumers are not treated as individuals but as a continuum. That is, we may consider that players approximately in state $x_t$ around time $t$ in reality buy or sell energy in bursts but that the population of order $O(N \d \, m(t,x_t))$ players, to which they belong, buy and sell electricity continuously on average.

\subsection{Mean Field Equilibrium}

Our interest now is to transpose the notion of own-state feedback NE into the corresponding notion of equilibrium in the mean field game, namely the own-state feedback mean field equilibrium (MFE). Based on Definition \ref{DefNE}, we state the following definition. 
\begin{definition}\label{DefMFE}
	The control $\alpha^\star \in \bar{\mathcal{A}}$ is a {\it mean field equilibrium in (own-state) feedback strategies} if, for all $\alpha \in \bar{\mathcal{A}}$ consistent with $m^\star$, it holds that
\begin{equation}
	J\left(\alpha^\star ; m^\star\right) \leq J_k\left(\alpha; m^\star \right),
\end{equation}
where $J\left(\cdot ; m^\star\right)$ denotes $J\left(\cdot\right)$ with $m$ replaced by $m^\star$ in its expression, $m^\star$ being the distribution induced by the mean field equilibrium $\alpha^\star$ for the dynamics \eqref{EqStateSDE} and for a given initial state distribution $m_0$.
%with associated distribution $m_t$, $m_t'$, the distributions of $x_t$, $x_t'$ induced by the controls $\alpha_t$, $\alpha_t'$, respectively, for a common initial individual state distribution $m_0$ and following the dynamics \eqref{EqStateSDE}, for all $t \in [0,1]$.
\end{definition}

%\red{A MFE $\alpha^{\star} \in \mathcal{A}$, if it exists, is an argument of the stochastic control problem
%\begin{align}
%	\label{eq:control_pb}
%	v(0,x_{0}) &= \inf_{\alpha \in \mathcal{A}} \left[J \left(\red{x_0,} \alpha \right) \right] \nonumber  \\
%	{\mathrm d} x_t &= \alpha_t {\mathrm d} t - g_t \left[{\mathrm d} t + \sigma_t {\mathrm d} W_t \right] + {\mathrm d} N_t,
%\end{align}
%with initial state $x_0$ and $m_0$. }

%\blue{Following Definition \ref{DefMFE}, a MFE can be written in terms of stochastic control problem as follows.}

Let us define the {\it value function} $v: [0,T] \times [0,1] \rightarrow \mathds{R}$, $(u,y) \rightarrow v(u,y)$, as follows,
\begin{align*}
	v(u,y) &= \inf_{\alpha \in \bar{\mathcal{A}} } \EE\left[ \displaystyle\int_u^T \left( \alpha_t p_t(m_t) + h_t(\alpha_{t}) + f_t(x_t)\right) {\mathrm d} t + \kappa (x_T)\right]
\end{align*}
where $x_t$ is any solution to \eqref{EqStateSDE} with $x_u=y$. 

According to \cite{GUE11}, an MFE $\alpha^\star$ for the game that generates a regular couple $(v,m)$ must be a solution to the following (backward) Hamilton-Jacobi-Bellman equation
\begin{align}
	\label{eq:HJB}
	\partial_t v(t,x) &= - \inf_{\alpha \in \mathds{R}} \left\{ \alpha \partial_x v(t,x) + \alpha p_t(m_t^{\star}) + h_t(\alpha_t) + f_t(x_t) \right\} \nonumber \\ 
	&+ g_t \partial_x v(t,x) - \frac12g_t^2\sigma_t^2\partial^2_{xx}v(t,x)
\end{align}
where $m(t,\cdot)^\star=m_t^{\star}$ is the solution of the following (forward) Fokker-Planck-Kolmogorov equation
\begin{align}
	\label{eq:FPK}
\partial_t m(t,x) &= - \partial_x \left[(\alpha_t^\star-g_t)m(t,x)\right] + \frac12 g_t^2\sigma_t^2 \partial^2_{xx}m(t,x)
\end{align}
for given $m(0,\cdot)$.

In the following, we assume the cost $h(\alpha_t)$ for control quadratic, i.e.
\begin{equation*}
	h_{t}(\alpha) = \frac12 H_t \alpha^2,
\end{equation*}
with $H_t > 0$ representing the unwillingness of the car owner to buy or sell energy at time $t$. This choice is seemingly non-natural as it implies that users are more willing to buy or sell small quantities rather than large quantities of energy. Nonetheless, under the mean field game formulation, this has to be understood as the fact that, on average, only a limited population of users at time $t$ is willing (or able) to buy energy. As such, intuitively, making the (psychological) cost of buying or selling energy larger for larger amounts of energy forces only part of the population to buy or sell. As for the particular choice of a quadratic cost rather than any other cost function, it is convenient for calculus mostly.

Under this assumption, solving
\begin{equation*}
	\inf_{\alpha \in \mathds{R}} \left\{ \alpha \partial_x v(t,x) + \alpha p_t(m_t^{\star}) + h_t(\alpha_t) + f_t(x_t) \right\}
\end{equation*}
for all $t$, it is immediate by convexity arguments to see that the optimal trajectory $\alpha^\star$ is explicitly given by
\begin{equation}
	\alpha_t^\star  = -\frac1{H_t}\left[\partial_x v(t,x) + p_t(m_t^{\star}) \right],
\end{equation}
possibly subject to some boundary conditions to ensure that $x_t\in[0,1]$ at all times. In the remainder of the article, we will assume this condition always met, so that at no time we will consider EV owners with completely full or completely empty batteries.

The HJB equation now becomes
\begin{align*}
	0&=\partial_t v(t,x) - \left(\frac1{H_t}\left[\partial_x v(t,x)+p_t(m_t^\star) \right] + g_t \right)\partial_x v(t,x) \\ 
	&- \frac{p_t(m_t^\star)}{H_t}\left[\partial_x v(t,x)+p_t(m_t^\star) \right] + f_t(x) \\ 
	&+ \frac1{2{H}_t}\left[\partial_x v(t,x)+p_t(m_t^\star) \right]^2 + \frac12\sigma_t^2 g_t^2 \partial^2_{xx}v(t,x),
\end{align*}
which can be simplified as
\begin{align*}
	\partial_t v(t,x) &= \frac1{2{H}_t}(\partial_x v(t,x)+p_t(m_t^\star))^2 + g_t \partial_x v(t,x)  \\ 
	&- {f_t(x)}  - \frac12\sigma_t^2g_t^2 \partial_{xx}^2v(t,x)
\end{align*}
and the FPK equation is
\begin{align*}
	\partial_t {m(t,x)} &=\left(\frac1{{H}_t}\left[\partial_x v^\star(t,x)+p_t({m(t,x)}\right] + g_t \right)\partial_x {m(t,x)} \\ 
	&+  \frac1{{H}_t}\partial^2_{xx} v^\star(t,x) {m(t,x)}  + \frac12g_t^2\sigma_t^2 \partial^2_{xx}{m(t,x)}.
\end{align*}
This defines the two fundamental differential equations to be solved, either explicitly or numerically, for determining the MFE. 

 In the next section, we improve the EV framework by turning the purely electrical vehicles into PHEV, introducing therefore the possibility for players to select between two alternative sources of energy.

\section{Plug-in Hybrid Vehicles}\label{SecPHEV}
\subsection{System Model}
In this section, we consider that vehicles in the set $\mathcal{K}$ are PHEV. A PHEV can operate both with an electrical energy source and an alternative energy source, for instance oil. The PHEV interacts with the electricity distribution grid by trading electricity with an elastic price, while trading oil at a fixed price (which is a natural assumption on a daily or even weekly basis). %, indeed, the eventual sales of oil made by PHEV are not considered. 
We describe the energy reserves of PHEV $k$ by the two-dimensional vector $\bm{z}_t^{(k)} =(z_{1, t}^{(k)},z_{2,t}^{(k)})^\trans \in [0,1]^2$, where $z_{1, t}$ is the amount of energy stored in the batteries and $z_{2, t}$ the level of the oil tank. We denote the provisioning rates of electricity and oil of PHEV $k$ by $\mu_{1,t}^{(k)} \in \mathds{R}$ and $\mu_{2,t}^{(k)} \in \mathds{R}$, respectively.
In addition, we denote $\beta^{(k)}: \mathds{R}_+ \times [0,1]^2 \rightarrow [0,1]$, $(t,\bm{z})\mapsto \beta^{(k)}(t,\bm{z})$, with $\bm{z} = \left( z_1, z_2 \right)$, the function that determines the relative proportion of energy drawn from the batteries of PHEV $k$ at time $t$. Typically, taking $\beta^{(k)}(t,\bm{z})= z_{1}/(z_{1}+z_{2})$ translates a policy where energy is consumed indistinctly of the energy source. Note that, depending on the typical distances covered by PHEV owners at time $t$ (e.g. weekdays against weekends), $\beta^{(k)}(t,\bm{z}^{{(k)}}_t)$ may explicitly depend on $t$. Alternatively, we may have considered $\beta^{(k)}(t,\bm{z}^{{(k)}}_t)$ an additional control variable which can be set optimally by the car owner depending on the status of the energy market. Nonetheless, for simplicity of analysis, we do not consider this scenario here.  We relate the variables $\bm{z}^{{(k)}}_t$, $\bm{\mu}^{{(k)}}_t$, and $\beta^{(k)}$ by the following state evolution dynamics
\begin{equation}\label{EqODEPHEV}
\frac{{\mathrm d}}{{\mathrm d} t }  \bm{z}^{{(k)}}_t = \begin{bmatrix} \mu^{{(k)}}_{1,t}  \\ \mu^{{(k)}}_{2,t} \end{bmatrix} - \begin{bmatrix} \beta(t, \bm{z}^{{(k)}}_t) \\ 1-\beta(t,\bm{z}^{{(k)}}_t) \end{bmatrix} g^{{(k)}}_t
\end{equation}
and, similar to previously, we consider only $\beta$ functions and $\bm{\mu}_t^{(k)}$ controls which are admissible, in the sense of their defining a unique solution $\bm{z}_t^{(k)}$ for each $t,k$.

We then define the cost of PHEV $k$ in the time window $[0,T]$ as
\begin{align}\label{EqCostFunctionPHEV}
L_k\left(\bm{\mu}^{(k)}, \bm{\mu}^{(-k)}\right) & = \int_0^T \left( \bm{r}_{t}\left(\bm{\mu}_{t}^{(1)}, \ldots, \bm{\mu}_{t}^{(K)} \right) + q_{t}^{(k)}(\bm{\mu}_t^{(k)})  + s_{t}^{(k)}(\bm{z}_t^{(k)}) \right) {\mathrm d} t 
%\\ & & 
+ \xi^{(k)}(\bm{z}_T^{(k)}),
\end{align}
for a given initial state $\bm{z}_{0} \in [ 0,1 ]^{2 K}$, i.e. the initial energy reserves of all PHEV, where $\bm{\mu} = (\bm{\mu}^{(1)}, \ldots, \bm{\mu}^{(K)})$, with $\bm{\mu}^{(k)} = \{\bm{\mu}_{t}^{(k)}=( \mu_{1,t}^{(k)}, \mu_{2,t}^{(k)}), 0 \leq t \leq T \}$ belonging to the set of admissible controls for the dynamics \eqref{EqODEPHEV}. 

Here, $\bm{r}_{t}: \mathds{R}^{2K} \rightarrow \mathds{R}^{2}$, $\bm{\mu}\mapsto \bm{r}_{t}(\bm{\mu}^{(1)}, \ldots, \bm{\mu}^{(K)}) = ( r_{1,t}(\mu_{1}^{(1)}, \ldots, \mu_{1}^{(K)}), r_{2,t}(\mu_{2}^{(1)}, \ldots, \mu_{2}^{(K)}) )$ evaluates the instantaneous prices $r_{1,t}$ of electricity and $r_{2,t}$ of oil, given the controls $\bm{\mu}^{(k)}=(\mu_{1}^{(k)},\mu_2^{(k)})$. In particular, we assume here that the price for oil is fixed, given by $r_{2,t}(\mu_{2}^{(1)}, \ldots, \mu_{2}^{(K)}) =r_{2}$.
Note that in this case the trajectory of the state $\bm{z} = \{ \bm{z}_t = ({z}_t^{(1)}, \ldots, {z}_t^{(N)}), 0 \leq t \leq T \}$ is determined by the initial state $\bm{z}_{0} = (z_{0}^{(1)}, \ldots, {z}_{0}^{(K)})$ and by the dynamics \eqref{EqODEPHEV}. We denote $\mathcal Z$ the set of state trajectories $\bm{z}$. 

The function $q_{t}^{(k)} :\mathds{R}^2 \rightarrow \mathds{R}$, $\bm{\mu}\mapsto q_{t}^{(k)}(\bm{\mu})$ evaluates the psychological cost of trading a quantity $\mu_1$ of electricity and a quantity $\mu_2$ of oil at time $t$, where $\bm{\mu}=(\mu_1,\mu_2)^\trans$. The function $s_{t}^{(k)}: [0, 1]^2 \rightarrow \mathds{R}$, $\bm{z}\mapsto s_t^{(k)}(\bm{z})$ denotes the cost for PHEV $k$ to be in state $\bm{z}=(z_1,z_2)$ at time $t$. Finally, $\xi^{(k)}: [0, 1]^2 \rightarrow \mathds{R}$, ${\bm{z}}\mapsto \xi^{(k)}(\bm{z})$ is the cost for PHEV $k$ to be in state $\bm{z}=(z_1,z_2)$ at time $T$.
These are analogous to the functions $h_{t}^{(k)}$, $f_{t}^{(k)}$, and $\kappa^{(k)}$ in \eqref{EqCostFunctionEV}, respectively.
 
In the following, we formulate the finite-number of players differential game.  
%%%%%%%%%%%%%%%%%%%%%%%%

\subsection{Classical Game Formulation}
The interaction between all PHEVs is modeled by a $K$-player continuous-time stochastic differential game of pre-specified fixed duration $T>0$. As for the case of EV, the aim of player $k$ is to determine the control trajectory $\bm{\mu}^{(k)} = \{\bm{\mu}_{t}^{(k)}, 0 \leq t \leq T \}$ such that its cost $L_k$ in \eqref{EqCostFunctionPHEV} is minimized given the initial conditions $\bm{z}_0$ and the control trajectories adopted by all the other players $\bm{\mu}^{(-k)}$. We denote the set of all admissible controls $\bm{\mu}^{(k)}$ of player $k$ over the time period $[0,T]$ by $\mathcal{U}_k$, and we denote $\mathcal U=\mathcal{U}_1\times \cdots\times \mathcal{U}_K$. At time $t$, the instantaneous control $\bm{\mu}^{(k)}$ is determined based on the information available to player $k$, which we denote by the information set $\eta_{t}^{(k)}$, as in the previous section. Here, the information set corresponds to the singleton $\eta_{t}^{(k)} = \lbrace \bm{z}^{(k)}_t \rbrace$.
Let us denote the  strategy of player $k$ by $\theta_{t}^{(k)}: \eta_t^{(k)} \rightarrow \mathcal{U}_k$, $\eta_t^{(k)} \rightarrow \theta_{t}^{(k)}( \eta_t^{(k)})$. As stated above, this strategy corresponds to the class of non-anticipative own-state feedback strategies, and we will write
\begin{equation}\label{EqAlphaDefPHEV}
\bm{\mu}_t^{(k)} =  \theta_{t}^{(k)}(\eta_t^{(k)}).
\end{equation}
The image of $\theta_{t}^{(k)}$, i.e. the set of own-state feedback controls, is denoted $\bar{\mathcal U}_k$ and we write $\bar{\mathcal U}=\bar{\mathcal U}_1\times \ldots\times \bar{\mathcal U}_K$.

\subsection{Mean Field Game Formulation}\label{SecMFGformulationPHEV}
In this section, we proceed similarly to Section \ref{SecMFGformulationEV}. We use here a finite-game counting measure $m^{K}$ of the form
\begin{equation}
	m^K(t, \bm{z}) =  \frac{1}{K}\sum_{k=1}^K \delta_{({z}_{1,t}^{(k)},{z}_{2,t}^{(k)})=(z_1,z_2)}, 
\end{equation}
with $\bm{z}=(z_1,z_2)^\trans$, and we assume asymptotic player indistinguishability to ensure that it admits a weak limiting distribution $m(t,\cdot)$ as $K\to\infty$. 
As previously, the individual state of each player is assumed to be a noisy version of the deterministic state trajectory in \eqref{EqODEPHEV} determined by the following SDE,
\begin{eqnarray}
\nonumber	
	{\mathrm d} \bm{z}_{t}  &= & \begin{bmatrix} \mu_{1,t}  \\ \mu_{2,t} \end{bmatrix} {\mathrm d} t - \begin{bmatrix} \beta(t,\bm{z}_{t} ) \\ 1-\beta (t,\bm{z}_{t} ) \end{bmatrix} g_t  \left(\begin{bmatrix} 1  \\ 1 \end{bmatrix} + \sigma_t {\mathrm d} \bm{W}_t \right)
	%\\ & &
	 + {\mathrm d} \bm{N}_t
\end{eqnarray} 
for a given initial state $\bm{z}_{0}$. In particular, $\bm{W}_t = \left(W_{1,t} , W_{2,t} \right)^{\trans}$ is a two-dimensional Brownian motion with independent components and $\bm{N}_t$ is the associated reflection vector.
Similar to the EV scenario, $\sigma_t$ determines the variance of the noise at time $t$. 
The analysis of the game now reduces to the analysis of the behavior of a single player. The cost function $L^{(k)}\triangleq L$, assumed identical to all players, reads
\begin{eqnarray}\label{EqCostFunctionPHEVinfty}
	L\left(\bm{\mu},m \right) & = & \EE \int_0^T \left( \bm{r}_{t}\left( m_t \right) + q_{t} (\bm{\mu}_t )  + s_{t} (\bm{z}_t) \right) {\mathrm d} t 
%\\ & & 
+ \xi(\bm{z}_T),
\end{eqnarray} 
where $m_t=m(t,\cdot)\in \mathcal{M}_t$ is the distribution of the state variable $\bm{z}_t$ and $\mathcal{M}_t$ is the set of distributions at time $t$. The initial state condition is $\bm{z}_{0} \in [0,1]^2$, a random variable with distribution $m_0$. The price for electricity is given by the function $r_{1,t}: \mathcal{M}_t \rightarrow \mathds{R}_+$, with
\begin{equation} 
	r_{1,t} (m_t) =  D(t,\cdot)^{-1} \left( g_t \int_{[0,1]^2} \beta(t,\bm{z})m_t(\bm{z}) {\mathrm d} \bm{z} + \frac{d}{dt} \int_{[0,1]^2}  z_1 {m_t(\bm{z})} {\mathrm d} \bm{z} \right)
\end{equation}
for $\bm{z}=(z_1,z_2)^\trans$ in the integrals.
The price for oil is constant, given by $r_{2,t} = r_{2}$.

The next section is dedicated to determining the MFE for this game.

\subsection{Mean Field Analysis}

Under the above game formulation, the optimal control problem which represents the equilibrium of the game formulates as
\begin{eqnarray}
\nonumber 
u(0,\bm{z}_{0}) & = & \inf_{\bm{\mu} \in \bar{\mathcal{U}}} L\left(\bm{\mu}, m_0 \right) \\
\label{EqSOPPHEV}	
	  {\mathrm d} \bm{z}_t & = & \begin{bmatrix} \mu_{1,t}  \\  \mu_{2,t}  \end{bmatrix} {\mathrm d} t - \begin{bmatrix} \beta(t,\bm{z}_t)\\ 1-\beta(t,\bm{z}_t) \end{bmatrix} g_t \left[{\mathrm d} t + \sigma_t {\mathrm d} \bm{W}_t \right] + {\mathrm d} \bm{N}_t.
\end{eqnarray}

We introduce the value function
\begin{equation}
	v(u,{\bm{y}}) =
	\inf_{\bm{\mu} \in \bar{\mathcal{U}}} \EE \left[\int_u^T \left( \bm{r}_{t}\left( m_t \right) + q_{t} (\bm{\mu}_t )  + s_{t} (\bm{z}_t) \right)  {\mathrm d} t + \xi(\bm{z}_T)\right]
\end{equation}
with {initial value} $\bm{z}_u=\bm{y}$.

As in the EV case, we consider the cost function as quadratic, that is,
\begin{equation*}
q_t(\bm{\mu}_t) = \frac{1}{2} {Q}_{1,t} (\mu_{1,t})^2 + \frac{1}{2} {Q}_{2,t} (\mu_{2,t})^2,
\end{equation*}
with $({Q}_{1,t},{Q}_{2,t}) \in \mathds{R}^2$.

The HJB equation, which provides a necessary condition for the existence of an MFE generating regular couples $(v,m)$, is here given by
\begin{eqnarray}
\nonumber
-\partial_t v(t,\bm{z})  & = & \inf_{({\mu}_{1,t},\mu_{2,t}) \in \mathds{R}^2} \left\lbrace \mu_{1,t} \, r_{1,t}\left( m_t^{\star} \right)  + \mu_{2,t} r_{2}  + q_t\left( \bm{\mu}_t \right)   
\right. \\ & & \nonumber \left.
+ \right(\mu_{1,t}  - g_t\beta(t,\bm{z}) \left)\partial_{{z_1}} v(t,\bm{z})  
\right. \\ & & \nonumber \left.
+ \left(\mu_{2,t} + g_t(\beta\left(t,\bm{z}\right) - 1)\right) \partial_{{z_2}} v(t,\bm{z}) \right\rbrace 
 \\ & & \nonumber
+ f_t(\bm{z})  + \frac12\sigma_t^2 \, g_t^2 \left[\right(\beta(t,\bm{z}) \left)^2 \partial^2_{{z_1}{z_1}}v(t,\bm{z})  
\right. \\ & & \nonumber \left.
+ 2\beta(t,\bm{z})(1-\beta(t,\bm{z}))\partial^2_{\bm{z}}v(t,\bm{z}) 
\right. \\ & & \label{EqHJBinPHEV}\left.
+ (1-\beta(t,\bm{z}))^2 \partial^2_{{z_2}{z_2}}v(t,\bm{z}) \right],
\end{eqnarray}
where ${m}^{\star}=\{m_t^\star,0\leq t\leq T\}$, $m_t^\star=m(t,\cdot)^\star$, is solution to the FPK equation
\begin{eqnarray}
\nonumber 	
	\partial_t m\left( t, \bm{z} \right) & = & 
	- \partial_{{z_1}} \left[(\mu_{1,t}^{\star}-\beta\left( t, \bm{z} \right)  g_t)  {m\left( t, \bm{z} \right)} \right] 
\\ \nonumber & &	
	- \partial_{{z_2}} \left[(\mu_{2,t}^{\star}+(\beta\left( t, \bm{z} \right) -1)g_t) {m\left( t, \bm{z} \right)}\right] 
\\ \nonumber & &	
	+ \frac12 g_t^2\sigma_t^2 \left[\beta\left( t, \bm{z} \right) ^2\partial^2_{{z_1}{z_1}} {m\left( t, \bm{z} \right)}+ (1-\beta\left( t, \bm{z} \right))^2 \partial^2_{{z_2}{z_2}}  {m\left( t, \bm{z} \right)}  
\right. \\ \nonumber & &	\left.	
	+2\beta\left( t, \bm{z} \right) (1-\beta\left( t, \bm{z} \right) )\partial^2_{{z_1}{z_2}} {m\left( t, \bm{z} \right)}\right],
\end{eqnarray}
with $\bm{\mu}_t^\star = ( \mu_{1,t}^\star, \mu_{2,t}^\star)\in\bar{\mathcal U}_t$ the cost minimizing ($\bm{z}_t$-adapted) feedback control, determined by
\begin{eqnarray}
	\mu_{1,t}^{\star} &=& -\frac1{{Q}_{1,t} } (r_{1,t}\left(m_t^\star \right)+\partial_{{z_1}}v(t,\bm{z}))\\
	\mu_{2,t}^{\star} &=& -\frac1{{Q}_{2,t} } \left(r_{2} + \partial_{{z_2}} v(t,\bm{z})\right).
\end{eqnarray}
 
Assuming $\sigma_t=0$, we obtain more compact forms. In particular, after substitution of the expression of $\bm{\mu}_t^\star$, the HJB equation becomes
\begin{eqnarray}\label{EqHJBphev}
	\partial_t v\left(t,\bm{z}\right) &=&  
	\frac1{2 {Q}_{1,t} }\left(\partial_{{z_1}}v\left(t,\bm{z}\right) +r_{1,t}(m_t^\star)\right)^2 
\nonumber \\ & &	
	+ \frac1{2 {Q}_{2,t} }\left(\partial_{{z_2}}v\left(t,\bm{z}\right) +r_{2}\right)^2 
\nonumber \\ & &	
	+ g_t\beta\left(t,\bm{z}\right)  \partial_{{z_1}}v\left(t,\bm{z}\right) 
\nonumber \\ & &	
	 + g_t (1-\beta\left(t,\bm{z}\right) )\partial_{{z_2}}v\left(t,\bm{z}\right)  - f_t\left(\bm{z}\right).
\end{eqnarray}
where $m_t^\star$ is solution to
\begin{eqnarray}\label{EqFPKphev}
	\nonumber
	\partial_t  {m\left( t, \bm{z} \right)}&=&
	\left[\frac1{{Q}_{1,t}}(\partial_{{z_1}{z_1}}v^\star) + \frac1{{Q}_{2,t}}(\partial_{{z_2}{z_2}}v^\star) + g_t\left[\partial_{{z_1}}\beta\left(t,\bm{z}\right) - \partial_{{z_2}}\beta\left(t,\bm{z}\right) \right]\right]   {m\left( t, \bm{z} \right)}
	 \\  \nonumber & &
	 + \left[\frac1{{Q}_{1,t}}(r_{1,t}({m\left( t, \bm{z} \right)}) + \partial_{{z_1}}v^\star)+\beta\left(t,\bm{z}\right) g_t\right] \partial_{{z_1}}  {m\left( t, \bm{z} \right)}  
	 \\  & &
	 + \left[\frac1{{Q}_{2,t}}(r_{2} + \partial_{{z_2}}v^\star)+(1-\beta\left(t,\bm{z}\right))g_t\right]\partial_{{z_2}}  {m\left( t, \bm{z} \right)} 
 \end{eqnarray}
 with $\bm{v}^\star$ the solution to \eqref{EqHJBphev}, which is our final expression. Note in particular that, for $\beta(t,\bm{z})=\frac{{z_1}}{{z_1}+{z_2}}$, $\bm{z}=(z_1,z_2)^\trans$, which we will use in Section \ref{SecSimulations}, we have that
\begin{equation}
	\partial_{{z_1}}\beta\left(t,\bm{z}\right) - \partial_{{z_2}}\beta\left(t,\bm{z}\right) = \frac1{{z_1}+{z_2}}.
\end{equation}

%%%%%%%%%%%%%%%%%%%%%%%%%%%%%%

\section{Simulations}
\label{SecSimulations}

In this section, we provide simulation results for the electrical vehicle schemes developed in Section \ref{SecPurelyEV} and Section \ref{SecPHEV}.

\subsection{EV analysis}
We first consider the scenario of Section \ref{SecPurelyEV}. We assume a realistic three-day scenario ($t=0$ at midnight the first day and $t=T=1$ seventy-two hours later) where players have an average consumption rate that depends on specific periods of the days. The scenario is typical of a Friday to Sunday energy consumption, with higher overall electricity consumption on Friday and different patterns of car usage on Friday than on Saturday and Sunday. Since it is difficult to provide a universal system parametrization, we will take arbitrary scalings in the energy consumption functions. 

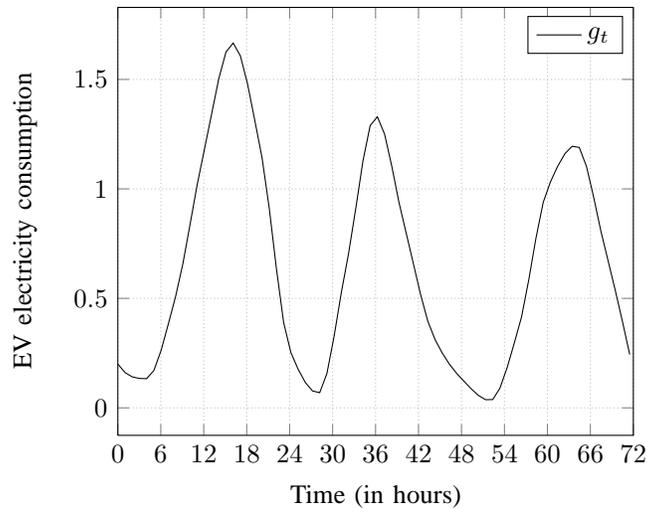
\begin{figure}
\centering
	\begin{tikzpicture}
		\begin{axis}[
			xlabel={Time (in hours)},
			ylabel={EV electricity consumption},
			grid=major,
			xmin=0,
			xmax=72,
			xtick={0,6,...,72},
			legend style={at={(0.98,0.98)}, anchor={north east}}, font=\small]
			\pgfplotsset{every major grid/.style={densely dotted}}
			\addlegendentry{{$g_t$}};
			%\addlegendentry{{$(n_1^*,n_2^*)$}};

			\addplot[] plot coordinates {
(0.000000,0.200000)(1.006993,0.161262)(2.013986,0.141493)(3.020979,0.134311)(4.027972,0.133364)(5.034965,0.170794)(6.041958,0.262081)(7.048951,0.382315)(8.055944,0.506711)(9.062937,0.654365)(10.069930,0.833975)(11.076923,1.017897)(12.083916,1.178808)(13.090909,1.340381)(14.097902,1.502778)(15.104895,1.624970)(16.111888,1.666004)(17.118881,1.607417)(18.125874,1.478173)(19.132867,1.312459)(20.139860,1.142896)(21.146853,0.912553)(22.153846,0.634904)(23.160839,0.389394)(24.167832,0.253036)(25.174825,0.177093)(26.181818,0.116185)(27.188811,0.077276)(28.195804,0.069400)(29.202797,0.156860)(30.209790,0.327428)(31.216783,0.526686)(32.223776,0.704183)(33.230769,0.910316)(34.237762,1.127739)(35.244755,1.289273)(36.251748,1.329497)(37.258741,1.249888)(38.265734,1.103356)(39.272727,0.937766)(40.279720,0.797394)(41.286713,0.657453)(42.293706,0.516563)(43.300699,0.395237)(44.307692,0.311967)(45.314685,0.250068)(46.321678,0.199068)(47.328671,0.157142)(48.335664,0.121949)(49.342657,0.087344)(50.349650,0.057022)(51.356643,0.037367)(52.363636,0.037815)(53.370629,0.090214)(54.377622,0.184049)(55.384615,0.297314)(56.391608,0.415859)(57.398601,0.585229)(58.405594,0.776250)(59.412587,0.938579)(60.419580,1.031689)(61.426573,1.103047)(62.433566,1.160372)(63.440559,1.194473)(64.447552,1.189471)(65.454545,1.102823)(66.461538,0.961881)(67.468531,0.806994)(68.475524,0.671651)(69.482517,0.535684)(70.489510,0.392833)(71.496503,0.243564)
			};
		\end{axis}
	\end{tikzpicture}
	\caption{Mean energy consumption $g_t$ of EV as a function of time.}
	\label{fig:g}
\end{figure}

\begin{figure}
\centering

	\begin{tikzpicture}
		\begin{axis}[
			xlabel={Time (in hours)},
			ylabel={Total electricity consumption},
			grid=major,
			xmin=0,
			xmax=72,
			xtick={0,6,...,72},
			legend style={at={(0.98,0.98)}, anchor={north east}}, font=\small]
			\pgfplotsset{every major grid/.style={densely dotted}}
			%\addlegendentry{{Approximation}};
			%\addlegendentry{{$(n_1^*,n_2^*)$}};

			\addplot[smooth,densely dashed] plot coordinates {
			(0.000000,1.253685)(1.006993,1.243615)(2.013986,1.233545)(3.020979,1.220924)(4.027972,1.213690)(5.034965,1.219472)(6.041958,1.234668)(7.048951,1.343771)(8.055944,1.456293)(9.062937,1.493544)(10.069930,1.514734)(11.076923,1.528278)(12.083916,1.533583)(13.090909,1.522325)(14.097902,1.513824)(15.104895,1.525253)(16.111888,1.533442)(17.118881,1.515651)(18.125874,1.491168)(19.132867,1.471028)(20.139860,1.451004)(21.146853,1.434319)(22.153846,1.406855)(23.160839,1.308193)(24.167832,1.226344)(25.174825,1.191170)(26.181818,1.171116)(27.188811,1.160819)(28.195804,1.151554)(29.202797,1.138774)(30.209790,1.135139)(31.216783,1.171730)(32.223776,1.226845)(33.230769,1.307692)(34.237762,1.360990)(35.244755,1.386404)(36.251748,1.393091)(37.258741,1.382827)(38.265734,1.370722)(39.272727,1.358013)(40.279720,1.354750)(41.286713,1.367868)(42.293706,1.372518)(43.300699,1.359311)(44.307692,1.354960)(45.314685,1.368250)(46.321678,1.370226)(47.328671,1.328897)(48.335664,1.274326)(49.342657,1.202170)(50.349650,1.168488)(51.356643,1.159116)(52.363636,1.149508)(53.370629,1.137023)(54.377622,1.138220)(55.384615,1.180609)(56.391608,1.238661)(57.398601,1.320610)(58.405594,1.365973)(59.412587,1.389152)(60.419580,1.392110)(61.426573,1.380592)(62.433566,1.368614)(63.440559,1.356377)(64.447552,1.356242)(65.454545,1.370034)(66.461538,1.370982)(67.468531,1.357172)(68.475524,1.356540)(69.482517,1.370361)(70.489510,1.367536)(71.496503,1.324666)
			};
			\addplot[smooth,red] plot coordinates {
			(0.000000,1.268781)(1.006993,1.261595)(2.013986,1.252740)(3.020979,1.241504)(4.027972,1.235076)(5.034965,1.240591)(6.041958,1.254282)(7.048951,1.352280)(8.055944,1.453714)(9.062937,1.486258)(10.069930,1.504184)(11.076923,1.515114)(12.083916,1.518430)(13.090909,1.507674)(14.097902,1.497085)(15.104895,1.506269)(16.111888,1.512205)(17.118881,1.495901)(18.125874,1.474097)(19.132867,1.456090)(20.139860,1.438260)(21.146853,1.423504)(22.153846,1.399781)(23.160839,1.311755)(24.167832,1.237819)(25.174825,1.206100)(26.181818,1.188657)(27.188811,1.179554)(28.195804,1.172208)(29.202797,1.160260)(30.209790,1.157259)(31.216783,1.189756)(32.223776,1.240548)(33.230769,1.312016)(34.237762,1.358155)(35.244755,1.378944)(36.251748,1.382700)(37.258741,1.371860)(38.265734,1.360579)(39.272727,1.348118)(40.279720,1.345141)(41.286713,1.357043)(42.293706,1.362143)(43.300699,1.351752)(44.307692,1.348321)(45.314685,1.361831)(46.321678,1.364878)(47.328671,1.327646)(48.335664,1.279345)(49.342657,1.214024)(50.349650,1.184132)(51.356643,1.177824)(52.363636,1.168977)(53.370629,1.157988)(54.377622,1.160552)(55.384615,1.199235)(56.391608,1.253056)(57.398601,1.323912)(58.405594,1.361759)(59.412587,1.378687)(60.419580,1.379398)(61.426573,1.366774)(62.433566,1.354633)(63.440559,1.342589)(64.447552,1.341084)(65.454545,1.354457)(66.461538,1.356135)(67.468531,1.346572)(68.475524,1.347736)(69.482517,1.361242)(70.489510,1.362332)(71.496503,1.324827)
			};
			\legend{{No EV regulation},{EV regulation}}
		\end{axis}
	\end{tikzpicture}
	\caption{Total electricity consumption with or without EV regulation as a function of the time $t$.}
	\label{fig:total_consumption}
\end{figure}
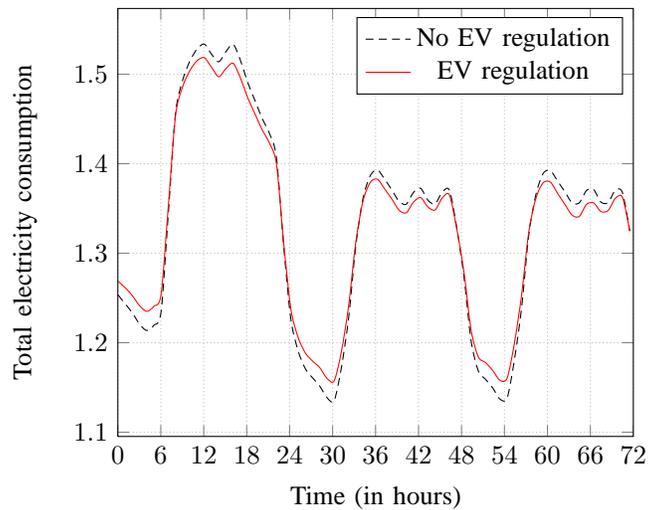

The car electricity consumption function $g_t$ is depicted in Figure \ref{fig:g}, where we see in particular that consumption is higher on Friday and with a peak around 5pm, while consumption is lower on weekend days with different peak times. The variance $\sigma_t^2$ on the consumption is taken equal to $0.01$ at all time, ensuring a standard deviation of the order of $10\%$. The demand function $D(t,p)$ is such that the price $p$ is a quadratic function of the total electricity demand from both electrical vehicles and other electricity services. Specifically, we take here 
$$p_t =\left(\left[g_t+\frac{d}{dt}\int x m(t,x)dx\right]^++d_t\right)^2$$
where $d_t$ stands for the demand of electricity in services other than electrical cars, with $[x]^+=\max(x,0)$. We therefore assume that this demand is deterministic and is not altered by the evolution of EV electricity price, which is a realistic assumption if the EV electricity market is independent of the outer electricity trading market. The function $d_t$ is depicted in dashed line in Figure \ref{fig:total_consumption}, up to a constant corresponding to the total average EV consumption; that is, the dashed line represents the total electricity consumption if EV consumption were distributed equally in time. For simplicity of understanding, we assume $h_t=30$ constant; that is, we do not consider that the car owners have any particular incentive to charge or discharge at some specific time periods.\footnote{Note that the determination of a correct $h_t$ is highly subjective and is better kept constant for the sake of interpretation.} We take $f(t,x)=(1-x)^2$ to impose consumers to keep a certain level of electricity in their batteries, and the boundary condition $\kappa(x)=(1-x)^2$ in order to avoid large sales at the last minute. The initial condition on $m(0,\cdot)$ is a triangle distribution $m_0$ centered at $0.5$ and with support $[0.3,0.7]$. The boundary conditions on $m$ and $v$ are such that $\partial_x m(0,\cdot)=\partial_x m(1,\cdot)=\partial_x v(0,\cdot)=\partial_x v(1,\cdot)=0$ in order to force the energy content to lie in $[0,1]$.

To solve the system of equations \eqref{eq:HJB}, \eqref{eq:FPK} in $(m,v)$, we proceed by solving sequentially the HJB and FPK equations using a simple fixed-point algorithm until convergence. We do not ensure here that this algorithm does converge, neither do we ensure that the solution obtained is the solution sought for. Using a finite difference method on a sampling of $144$ points in the time axis (every 30min) and of $100$ points in the battery level axis, the above scheme leads to the distribution evolution ${\bm m}^\star$ depicted in Figure \ref{fig:m}. A few observations can be already made from this figure. We easily observe daily sequences of increases and decreases of the average battery levels. We see in particular that during nighttime, the battery levels increase, indicating that energy is purchased in nighttime and consumed during daytime. It is interesting to note that, due to the small variance $\sigma_t^2$ that was chosen, the overall tendency is for $m^\star(t,\cdot)$ to concentrate into a single mass when $t\to 1$. This is a usual phenomenon which determines the steady state if time were to continue with constant values for all time-dependent system parameters.

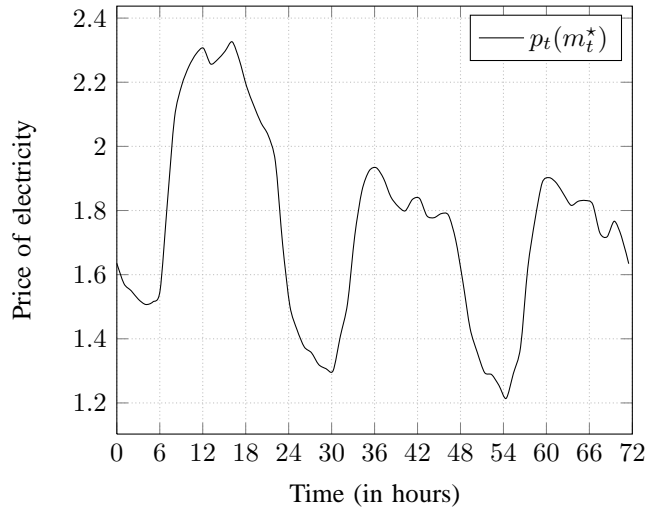
\begin{figure}
\centering

	\begin{tikzpicture}
		\begin{axis}[
			xlabel={Time (in hours)},
			ylabel={Price of electricity},
			grid=major,
			xmin=0,
			xmax=72,
			xtick={0,6,...,72},
			legend style={at={(0.98,0.98)}, anchor={north east}}, font=\small]
			\pgfplotsset{every major grid/.style={densely dotted}}
			\addlegendentry{{$p_t(m_t^\star)$}};
			%\addlegendentry{{Approximation}};
			%\addlegendentry{{$(n_1^*,n_2^*)$}};

			\addplot[smooth] plot coordinates {
(0.000000,1.635158)(1.006993,1.571481)(2.013986,1.549689)(3.020979,1.522927)(4.027972,1.507003)(5.034965,1.515006)(6.041958,1.550441)(7.048951,1.821031)(8.055944,2.088465)(9.062937,2.190351)(10.069930,2.249661)(11.076923,2.288590)(12.083916,2.306449)(13.090909,2.257079)(14.097902,2.271271)(15.104895,2.296399)(16.111888,2.326118)(17.118881,2.268729)(18.125874,2.186993)(19.132867,2.127032)(20.139860,2.073420)(21.146853,2.034094)(22.153846,1.951071)(23.160839,1.693919)(24.167832,1.503006)(25.174825,1.427805)(26.181818,1.374560)(27.188811,1.356484)(28.195804,1.319614)(29.202797,1.306997)(30.209790,1.300530)(31.216783,1.407811)(32.223776,1.510234)(33.230769,1.721330)(34.237762,1.861756)(35.244755,1.923668)(36.251748,1.933160)(37.258741,1.899768)(38.265734,1.842768)(39.272727,1.813198)(40.279720,1.798865)(41.286713,1.834618)(42.293706,1.837338)(43.300699,1.783588)(44.307692,1.777610)(45.314685,1.790252)(46.321678,1.785350)(47.328671,1.708167)(48.335664,1.577477)(49.342657,1.432133)(50.349650,1.357193)(51.356643,1.294911)(52.363636,1.288079)(53.370629,1.254652)(54.377622,1.214118)(55.384615,1.291779)(56.391608,1.374553)(57.398601,1.620823)(58.405594,1.771009)(59.412587,1.886875)(60.419580,1.902355)(61.426573,1.884959)(62.433566,1.849851)(63.440559,1.816245)(64.447552,1.828965)(65.454545,1.831168)(66.461538,1.819541)(67.468531,1.730005)(68.475524,1.718036)(69.482517,1.766653)(70.489510,1.718417)(71.496503,1.634067)
			};
		\end{axis}
	\end{tikzpicture}
	\caption{Evolution of the price $p_t({m}_t^\star)$ as a function of the time $t$.}
	\label{fig:price}
\end{figure}

From the expression of ${\bm m}^\star$, ${\bm v}^\star$, and the equations derived in Section \ref{SecPurelyEV}, it is now possible to obtain much information about the system. In particular, it is interesting to follow the electricity bought or sold by electrical vehicles at all time, that is the quantity
$$g_t+\frac{d}{dt}\int xm^\star(t,x)dx$$
or the overall electricity consumption in the market given by 
$$g_t+\frac{d}{dt}\int xm^\star(t,x)dx+d_t$$
and the price $p_t({m}_t^\star)$ defined here as
$$p_t({m}_t^\star) =\left(\left[g_t+\frac{d}{dt}\int xm^\star(t,x)dx\right]^++d_t\right)^2.$$
This is depicted in Figure \ref{fig:EV_consumption}, Figure \ref{fig:total_consumption} and in Figure \ref{fig:price}, respectively. 

We see first in Figure \ref{fig:EV_consumption} that the peaks of electricity bought by electrical vehicles take place during the night where the overall demand is low, while they are at their lowest during peak demand periods. This is a natural outcome of the fact that prices are high during peak demand periods. However, we also see that the difference of amplitude between lowest and highest purchases is not large. This is due to the fact that, while prices are high in peak demand periods, the EV owners still have a strong incentive not to find their batteries empty, driving them to keep buying electricity at peak periods. This behaviour can be hindered by relaxing the constraint $f(t,x)$. 

Of more interest is Figure \ref{fig:total_consumption}, where the differences between electricity consumption with or without incentives on EV behaviour is presented. This figure depicts in dashed line the overall energy consumption if the EV purchases were equally distributed in the three-day period (that is, with no incentive), and in plain line the overall consumption under our current assumptions. It is seen here that the price incentives on electricity purchases produce a much expected peak demand reduction in the critical day periods, and a simultaneous increase of consumption during low consumption periods. Note importantly that our analysis does not consider changes in $d_t$ when the price for electricity changes; only the part of electricity reserved for EV drives prices which in turn drive the EV behaviour, which is a natural assumption if different price conditions are applied to EV and other services. The price evolution is depicted in Figure \ref{fig:price}, where it is seen in this setting that the price is mostly driven by the function $d_t$.

%\begin{figure}
%
%	\begin{tikzpicture}
%		\begin{axis}[
%			xlabel={Time},
%			ylabel={Energy consumption},
%			grid=major,
%			xmin=0,
%			xmax=1,
%			xtick={0,0.2,0.4,0.6,0.8,1},
%			%ytick={0,0.2,0.4,0.6,0.8,1},
%			legend style={at={(0.98,0.98)}, anchor={north east}}, font=\small]
%			\pgfplotsset{every major grid/.style={densely dotted}}
%			\addlegendentry{{$g_t$}};
%			%\addlegendentry{{$(n_1^*,n_2^*)$}};
%
%			\addplot[] plot coordinates {
%			(0,0.2)(0.1,0.2)(0.3,1)(0.5,0.2)(0.6,0.2)(0.7,2.2)(0.8,0.2)(1,0.2)
%			};
%		\end{axis}
%	\end{tikzpicture}
%	\caption{Mean energy consumption $g_t$ of EV as a function of time.}
%	\label{fig:g}
%\end{figure}

\begin{figure}
\centering
	\begin{tikzpicture}
		\begin{axis}[
			view={0}{90},
			xlabel={Time $t$ (in hours)},
			ylabel={Battery level $x$},
			zlabel={Density},
			grid=none,
			xtick={0,6,...,72},
			%ytick={0,0.2,...,1},
			%ztick={0,0.2,...,1},
			%colormap={bw}{gray(0cm)=(1); gray(1cm)=(0)},
			%blackwhite,
			legend style={at={(0.98,0.98)}, anchor={north east}}, font=\small]
			\pgfplotsset{every major grid/.style={densely dotted}}
			\addlegendentry{{$m^\star(t,x)$}};
			%\addlegendentry{{Approximation}};
			%\addlegendentry{{$(n_1^*,n_2^*)$}};

			%\addplot3[surf, shader=interp] file {m_PHEV.txt};
			\addplot3[surf] file {m_PHEV2.txt};
			%\addplot3+[only marks,mark=+, mark size=2, mark options={color=black}] file {datadet40.txt};
			%\addplot3+[only marks,mark=x, mark size=4, mark options={color=red}, line width=1.5] coordinates {(9,1,11.82)};
		\end{axis}
	\end{tikzpicture}
	\caption{Density solution $m^\star(t,x)$ as a function of the time $t$ and the battery level $x$.}
	\label{fig:m}

\end{figure}
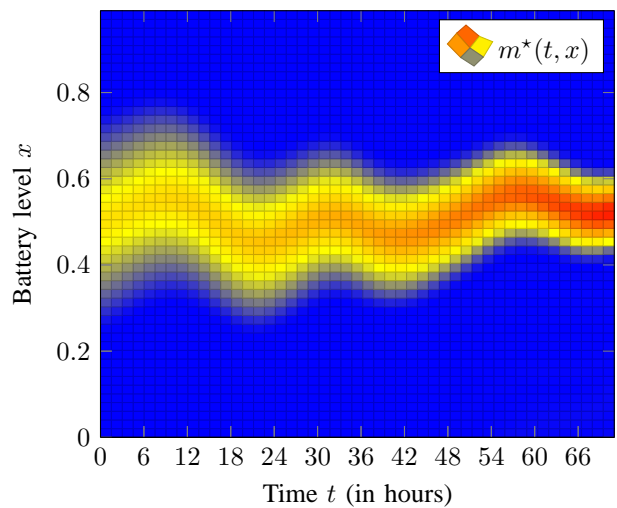

\begin{figure}
\centering

	\begin{tikzpicture}
		\begin{axis}[
			xlabel={Time (in hours)},
			ylabel={EV purchases},
			grid=major,
			xmin=0,
			xmax=72,
			xtick={0,6,...,72},
			legend style={at={(0.98,0.02)}, anchor={south east}}, font=\small]
			\pgfplotsset{every major grid/.style={densely dotted}}
			\addlegendentry{{$g_t+\frac{d}{dt}\int xm^\star(t,x)dx$}};
			%\addlegendentry{{$(n_1^*,n_2^*)$}};

			\addplot[smooth] plot coordinates {
			(0.000000,0.708781)(1.006993,0.711664)(2.013986,0.712881)(3.020979,0.714265)(4.027972,0.715071)(5.034965,0.714804)(6.041958,0.713299)(7.048951,0.702194)(8.055944,0.691106)(9.062937,0.686400)(10.069930,0.683136)(11.076923,0.680521)(12.083916,0.678533)(13.090909,0.679034)(14.097902,0.676946)(15.104895,0.674701)(16.111888,0.672449)(17.118881,0.673935)(18.125874,0.676615)(19.132867,0.678747)(20.139860,0.680941)(21.146853,0.682870)(22.153846,0.686611)(23.160839,0.697248)(24.167832,0.705160)(25.174825,0.708615)(26.181818,0.711226)(27.188811,0.712420)(28.195804,0.714339)(29.202797,0.715171)(30.209790,0.715806)(31.216783,0.711711)(32.223776,0.707388)(33.230769,0.698009)(34.237762,0.690850)(35.244755,0.686226)(36.251748,0.683293)(37.258741,0.682718)(38.265734,0.683543)(39.272727,0.683790)(40.279720,0.684077)(41.286713,0.682860)(42.293706,0.683311)(43.300699,0.686127)(44.307692,0.687046)(45.314685,0.687266)(46.321678,0.688336)(47.328671,0.692434)(48.335664,0.698705)(49.342657,0.705540)(50.349650,0.709329)(51.356643,0.712393)(52.363636,0.713155)(53.370629,0.714651)(54.377622,0.716018)(55.384615,0.712312)(56.391608,0.708080)(57.398601,0.696987)(58.405594,0.689472)(59.412587,0.683221)(60.419580,0.680974)(61.426573,0.679867)(62.433566,0.679705)(63.440559,0.679897)(64.447552,0.678528)(65.454545,0.678108)(66.461538,0.678838)(67.468531,0.683086)(68.475524,0.684882)(69.482517,0.684566)(70.489510,0.688482)(71.496503,0.693846)
			};
		\end{axis}
	\end{tikzpicture}
	\caption{Electricity purchased by EV as a function of the time $t$.}
	\label{fig:EV_consumption}
\end{figure}
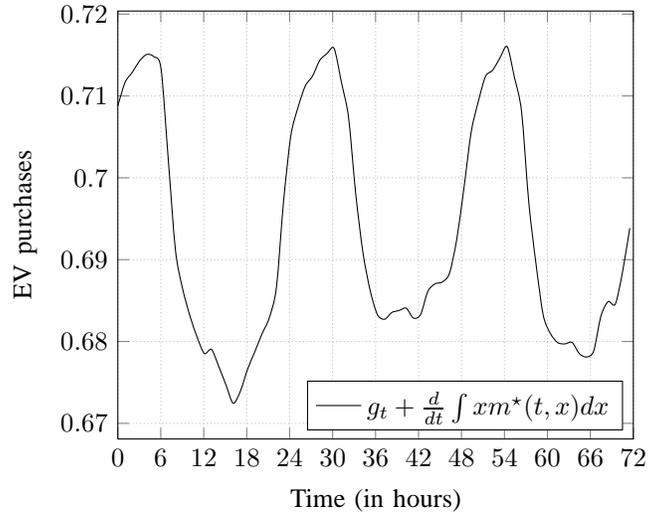

\subsection{PHEV analysis}
In this second section, we wish to analyze the behavior of hybrid vehicles as described in Section \ref{SecPHEV}. Since solving three-dimensional differential equations is time-consuming, we only provide results for the time scale discretized in $12$ samples and for the ``spatial'' scales discretized both in $16$ samples. For each differential equation, the resolution is performed by iterating the resolution of the two-dimensional differential equations along time and electricity scales for each fixed oil tank level, and time and oil scales for each fixed battery level. Then the system of HJB and FPK differential equations is solved by further iterating a fixed point algorithm as in the previous section. For simplicity of interpretation, we consider here a time-independent scenario where both $g_t=0.2$ and $(q_{1,t},q_{2,t})=(125,125)$ are constant with time.\footnote{Such a large value for the entries of $h_t$ is motivated by faster algorithm convergence reasons, although it inhibits as a counterpart fast variations of $m$ along time.} We take the electricity price to be $r_{1,t}=\left(D(t,r_{1,t})\right)^++0.5$, where now the demand is solely due to the electricity being bought by PHEVs; that is, we do not consider other sources of electricity consumption in order to focus on the oil/electricity interaction solely. The oil price is set to $r_{2,t}=r_2=0.7$. This is a natural choice as it is expected that an approximate quantity $g_t=0.2$ will be asked for at any time to cover for the energy consumed, hence a price for electricity $r_{1,t}\simeq 0.7$. We impose a constraint $s_t({\bm z})=20(2-z_1-z_2)^2$, where ${\bm z}=(z_1,z_2)^\trans$. The relative consumption $\beta$ of oil and electricity is proportional to the total quantity of energy, that is $\beta(t,{\bm z})=z_1/(z_1+z_2)$ and therefore $1-\beta(t,{\bm z})=z_2/(z_1+z_2)$. We take $\sigma_t=0$ for simplicity. The boundary constraints are identical to those in the previous section. As for the terminal constraint on $v$, it imposes that $v(T,{\bm z})=\xi({\bm z})=10(2-(z_1+z_2))^2$.

We consider the scenario where $m(0,\cdot)$ is a (properly truncated and scaled) Gaussian distribution with mean $(0.4,0.6)^\trans$ and covariance $0.02 I_2$, with $I_2$ the $2\times 2$ identity matrix. That is, we assume that, initially, most vehicles have more oil than electricity. This is depicted in Figure \ref{fig:m1}. We then let the system evolve freely under the above set of constraints. It is natural to guess that the overall behavior is a decrease of either or both quantities of oil and electricity to zero if the prices are too high, or an increase of either or both quantities to one, if the prices are more reasonable. What is interesting to observe is the trajectory jointly followed by the players. The resulting final distribution $m^\star(T,\cdot)$ is depicted in Figure \ref{fig:m2}. What we observe in the aforementioned conditions is that the initial distribution has shifted towards an increase of both electricity and oil levels, with a stronger increase of the mean battery level. Another observation is that the distribution tends to stretch along the $z_1=z_2$ diagonal in the figure, translating the fact that oil and electricity are seen almost as equivalent goods due to the loosely constraining energy cost policy. 

Among the different further analyses, in Figure \ref{fig:alpha}, we consider a section of the distribution of the optimal transaction policy $\mu_{1,t}^\star$ and $\mu_{2,t}^\star$ at time $t=0^+$, for $z_{2,t}=0.5$ and $z_{2,t}=0.9$ (we remind that both $\mu_{1,t}^\star$ and $\mu_{2,t}^\star$ are functions of $t$, $z_{1,t}$ and $z_{2,t}$). That is, we observe the initial behavior of players with half-filled oil tanks and almost completely filled oil tanks. It is seen that, for users with a very low level of electricity, buying electricity is an appealing choice. This can be interpreted by the fact that, as few players are in strong need for energy, it is possible to acquire a large quantity of electricity at a reasonable price. Those players with low reserves of electricity are the main beneficiaries. For users with already a reasonable level of electricity though, electricity and oil are seen as equivalent goods. As a matter of fact, our results also show that, at time $t=0^+$, the price of electricity equals $r_{1,t}=0.706\simeq r_{2}$. That is, the players with low electricity levels draw as much of the electricity overhead (compared to oil) as is needed to reach an equilibrium price with oil. Now, it is also observed that, for users with large quantities of oil, electricity becomes a compelling purchase in order to further increase the total quantity of energy (since $f$ imposes $z_{1,t}+z_{2,t}$ to be close to $2$), hence a larger incentive for buying electricity when the battery level is not large. When both battery and tank levels are alike, we see that the quantity of electricity purchased is the same as the quantity of oil purchased.

\begin{figure}
\centering

	\begin{tikzpicture}
		\begin{axis}[
			view={0}{90},
			xlabel={Battery level},
			ylabel={Oil tank level},
			zlabel={Density},
			grid=none,
			xtick={0,0.2,0.4,0.6,0.8,1},
			ytick={0,0.2,0.4,0.6,0.8,1},
			ztick={0,1,2,3,4,5},
			zmin=0,
			%colormap/greenyellow,
			%colormap={bw}{gray(0cm)=(1); gray(1cm)=(0)},
			legend style={at={(0.98,0.98)}, anchor={north east}}, font=\small]
			\pgfplotsset{every major grid/.style={densely dotted}}
			\addlegendentry{{$m^\star(0,\cdot)$}};
			%\addlegendentry{{Approximation}};
			%\addlegendentry{{$(n_1^*,n_2^*)$}};

			%\addplot3[surf, shader=interp] file {m_PHEV1.txt};
			\addplot3[surf] file {m_PHEV1.txt};
			%\addplot3+[only marks,mark=+, mark size=2, mark options={color=black}] file {datadet40.txt};
			%\addplot3+[only marks,mark=x, mark size=4, mark options={color=red}, line width=1.5] coordinates {(9,1,11.82)};
		\end{axis}
	\end{tikzpicture}
	\caption{Initial distribution $m(0,\cdot)$ at time $t=0$, as a function of both levels of battery and oil tank.}
	\label{fig:m1}

\end{figure}
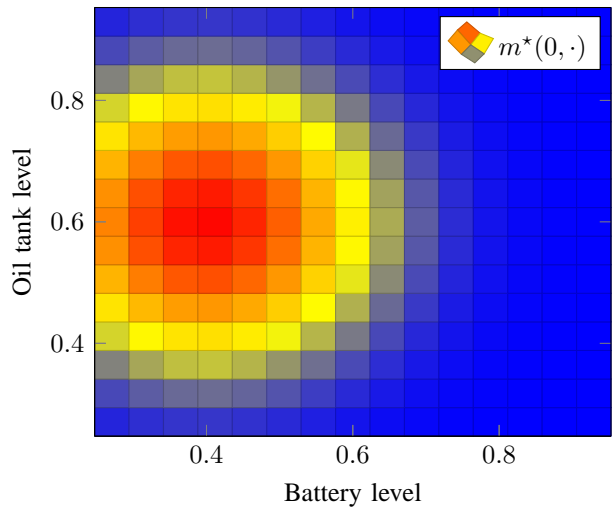

\begin{figure}
\centering

	\begin{tikzpicture}
		\begin{axis}[
			view={0}{90},
			xlabel={Battery level},
			ylabel={Oil tank level},
			zlabel={Density},
			grid=none,
			xtick={0,0.2,0.4,0.6,0.8,1},
			ytick={0,0.2,0.4,0.6,0.8,1},
			ztick={0,1,2,3,4,5},
			zmin=0,
			%colormap/greenyellow,
			%colormap={bw}{gray(0cm)=(1); gray(1cm)=(0)},
			legend style={at={(0.98,0.98)}, anchor={north east}}, font=\small]
			\pgfplotsset{every major grid/.style={densely dotted}}
			\addlegendentry{{$m^\star(T,\cdot)$}};
			%\addlegendentry{{Approximation}};
			%\addlegendentry{{$(n_1^*,n_2^*)$}};

			%\addplot3[surf, shader=interp] file {m_PHEVend.txt};
			\addplot3[surf] file {m_PHEVend.txt};
			%\addplot3+[only marks,mark=+, mark size=2, mark options={color=black}] file {datadet40.txt};
			%\addplot3+[only marks,mark=x, mark size=4, mark options={color=red}, line width=1.5] coordinates {(9,1,11.82)};
		\end{axis}
	\end{tikzpicture}
	\caption{Final distribution $m(T,\cdot)$ as a function of both levels of battery and oil tank.}
	\label{fig:m2}

\end{figure}
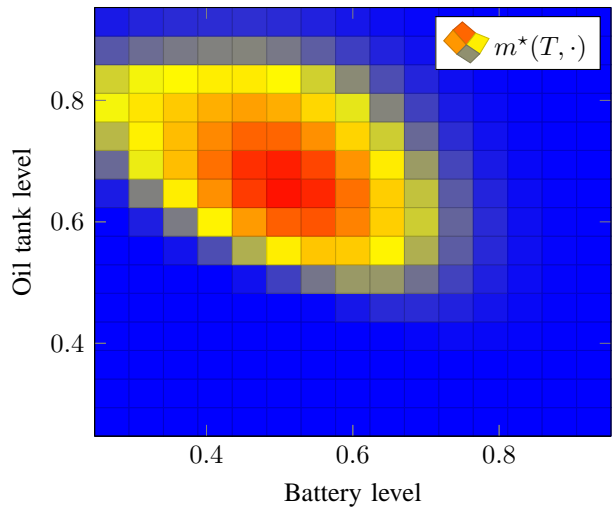
\begin{figure}
\centering

	\begin{tikzpicture}
		\begin{axis}[
			xlabel={Battery level},
			ylabel={Rate of energy purchase},
			grid=major,
			xtick={0,0.2,0.4,0.6,0.8,1},
			legend style={at={(0.98,0.98)}, anchor={north east}}, font=\small]
			\pgfplotsset{every major grid/.style={densely dotted}}
			\addlegendentry{{$\mu_{1,t}^{\star}$, $z_{2,t}=0.5$}};
			\addlegendentry{{$\mu_{2,t}^{\star}$, $z_{2,t}=0.5$}};
			\addlegendentry{{$\mu_{1,t}^{\star}$, $z_{2,t}=0.9$}};
			\addlegendentry{{$\mu_{2,t}^{\star}$, $z_{2,t}=0.9$}};
			%\addlegendentry{{$(n_1^*,n_2^*)$}};

			\addplot[smooth] plot coordinates {
			(0.247059,0.552117)(0.294118,0.478781)(0.341176,0.391320)(0.388235,0.385420)(0.435294,0.366173)(0.482353,0.348135)(0.529412,0.343002)(0.576471,0.322531)(0.623529,0.309528)(0.670588,0.294982)(0.717647,0.281665)(0.764706,0.267595)(0.811765,0.251160)(0.858824,0.228176)(0.905882,0.215963)(0.952941,0.218493)
			};
			\addplot[smooth,dashed] plot coordinates {
			(0.247059,0.426811)(0.294118,0.406649)(0.341176,0.394523)(0.388235,0.382492)(0.435294,0.367601)(0.482353,0.354346)(0.529412,0.338162)(0.576471,0.322991)(0.623529,0.308349)(0.670588,0.294514)(0.717647,0.280264)(0.764706,0.265224)(0.811765,0.250435)(0.858824,0.236600)(0.905882,0.225645)(0.952941,0.211640)
			};
			\addplot[smooth,blue,mark=x] plot coordinates {
(0.247059,0.383891)(0.294118,0.335436)(0.341176,0.264039)(0.388235,0.247298)(0.435294,0.237201)(0.482353,0.218564)(0.529412,0.207227)(0.576471,0.190998)(0.623529,0.176542)(0.670588,0.162048)(0.717647,0.147452)(0.764706,0.132718)(0.811765,0.117976)(0.858824,0.105096)(0.905882,0.091586)(0.952941,0.083624)
			};
			\addplot[smooth,blue,mark=x,dashed] plot coordinates {
(0.247059,0.241751)(0.294118,0.229063)(0.341176,0.219289)(0.388235,0.208563)(0.435294,0.196927)(0.482353,0.186655)(0.529412,0.175172)(0.576471,0.163870)(0.623529,0.152431)(0.670588,0.140788)(0.717647,0.128961)(0.764706,0.116893)(0.811765,0.104729)(0.858824,0.092211)(0.905882,0.080274)(0.952941,0.066695)
			};
		\end{axis}
	\end{tikzpicture}
	\caption{Optimal transactions at time $t=0^+$ for players with different oil and battery levels.}
	\label{fig:alpha}
\end{figure}
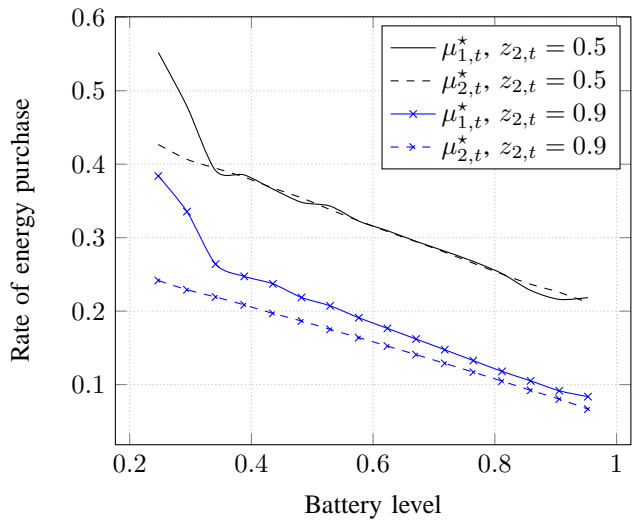

%\begin{figure}
%
%	\begin{tikzpicture}
%		\begin{axis}[
%			xlabel={Time},
%			ylabel={Prices},
%			grid=major,
%			xmin=0,
%			xmax=1,
%			ymin=0.5,
%			ymax = 0.8,
%			xtick={0,0.2,0.4,0.6,0.8,1},
%			%ytick={0,0.2,0.4,0.6,0.8,1},
%			legend style={at={(0.98,0.98)}, anchor={north east}}, font=\small]
%			\pgfplotsset{every major grid/.style={densely dotted}}
%			\addlegendentry{{$p_t^X(m^\star)$}};
%			\addlegendentry{{$p_t^Y$}};
%			%\addlegendentry{{$(n_1^*,n_2^*)$}};
%
%			\addplot[smooth] plot coordinates {
%			(0.000000,0.706000)(0.090909,0.701900)(0.181818,0.692900)(0.272727,0.682900)(0.363636,0.672100)(0.454545,0.658800)(0.545455,0.646800)(0.636364,0.635500)(0.727273,0.626800)(0.818182,0.618600)(0.909091,0.608000)(1.000000,0.603500)
%			};
%			\addplot[smooth,dashed] plot coordinates {
%			(0,0.7)(1,0.7)
%			};
%		\end{axis}
%	\end{tikzpicture}
%	\label{fig:price_PHEV}
%
%\end{figure}

Obviously, from the very generic settings of both EV and PHEV problems, many more scenarios can be carried out so to evaluate the actual impact of the EV and PHEV on realistic smart grid scenarios. The simulations above and their interpretations only provide a framework of fully rational vehicle owner's behavior, which needs be reported to real-life conditions with extreme care.

 \section{Conclusion}
\label{SecConclusion}

In this article, we proposed a game theoretical framework to model the behavior of electrical vehicle and hybrid electricity-oil vehicle owners aiming at selfishly minimizing their operating cost. As the number of selfish players is large, and players are assumed alike, we then turned the problem into a mean field game, for which we obtain the fundamental differential equations describing the mean field equilibrium of the game. Using numerical methods, we drew conclusions which give new insights on the way to optimize the electrical vehicle penetration in the future smart grid. 

\bibliography{./PHEV}

\begin{thebibliography}{10}
\providecommand{\url}[1]{#1}
\csname url@samestyle\endcsname
\providecommand{\newblock}{\relax}
\providecommand{\bibinfo}[2]{#2}
\providecommand{\BIBentrySTDinterwordspacing}{\spaceskip=0pt\relax}
\providecommand{\BIBentryALTinterwordstretchfactor}{4}
\providecommand{\BIBentryALTinterwordspacing}{\spaceskip=\fontdimen2\font plus
\BIBentryALTinterwordstretchfactor\fontdimen3\font minus
  \fontdimen4\font\relax}
\providecommand{\BIBforeignlanguage}[2]{{%
\expandafter\ifx\csname l@#1\endcsname\relax
\typeout{** WARNING: IEEEtran.bst: No hyphenation pattern has been}%
\typeout{** loaded for the language `#1'. Using the pattern for}%
\typeout{** the default language instead.}%
\else
\language=\csname l@#1\endcsname
\fi
#2}}
\providecommand{\BIBdecl}{\relax}
\BIBdecl

\bibitem{Simpson-2006}
A.~Simpson, ``Cost-benefit analysis of plug-in hybrid electric vehicle
  technology,'' in \emph{22nd International Battery, Hybrid and Fuel Cell
  Electric Vehicle Symposium}, Yokohama, Japan, Oct. 2006.

\bibitem{Bradley-2009}
T.~H. Bradley and A.~A. Frank, ``Design, demonstrations and sustainability
  impact assessments for plug-in hybrid electric vehicles,'' \emph{Elsevier
  Renewable and Sustainable Energy Reviews}, vol.~13, pp. 115--128, Jan. 2009.

\bibitem{KIN07}
M.~Kintner-Meyer, K.~Schneider, and R.~Pratt, ``Impacts assessment of plug-in
  hybrid vehicles on electric utilities and regional us power grids. part i:
  Technical analysis,'' \emph{Pacific Northwest National Laboratory}, 2007.

\bibitem{Pan-2010}
F.~Pan, R.~Bent, A.~Berscheid, and D.~Izraelevitz, ``Locating {PHEV} exchange
  stations in {V2g},'' in \emph{International Conference on Smart Grid
  Communications}, Gaithersburg, MD, USA, Oct. 2010, Oct. 2010.

\bibitem{Shao-2009}
S.~Shao, M.~Pipattanasomporn, and S.~Rahman, ``Challenges of {PHEV} penetration
  to the residential distribution network,'' \emph{IEEE Power and Energy
  Society (PES) General Meeting}, pp. 1--8, Jul. 2009.

\bibitem{Sojoudi-2011}
S.~Sojoudi and S.~H. Low, ``Optimal charging of plug-in hybrid electric
  vehicles in smart grids,'' in \emph{EEE Power and Energy Society (PES)
  General Meeting}, Gaithersburg, MD, USA, Oct. 2010, Oct. 2011.

\bibitem{Turton-2008}
H.~Turton and F.~Moura, ``Vehicle-to-grid systems for sustainable development:
  An integrated energy analysis,'' \emph{Technological Forecasting and Social
  Change}, vol.~75, pp. 1091--1108, Oct. 2008.

\bibitem{Cournot-1929}
A.~Cournot, ``Researches into the mathematical principles of the theory of
  wealth,'' \emph{Trans. N.T. Bacon, New York: Macmillan}, 1929.

\bibitem{Basar-Book}
T.~Ba\c{s}ar and G.~J. Olsder, \emph{Dynamic Noncooperative Game Theory}.\hskip
  1em plus 0.5em minus 0.4em\relax Philadelphia, PA, USA: SIAM Series in
  Classics in Applied Mathematics.

\bibitem{Saad-2011}
W.~Saad, Z.~Han, H.~V. Poor, and T.~Ba\c{s}ar, ``A noncooperative game for
  double auction-based energy trading between {PHEV}s and distribution grids,''
  in \emph{The 2nd IEEE International Conference on Smart Grid Communications
  (SmartGridComm)}, Brussels, Belgium, Oct. 2011.

\bibitem{Aumann-1964}
R.~Aumann, ``Markets with a continuum of trackers,'' \emph{Econometrica},
  vol.~32, pp. 39--50, 1964.

\bibitem{Carmona-2004}
G.~Carmona, ``Nash equilibria of games with a continuum of players,''
  \emph{Reprint}, 2004.

\bibitem{LAS06}
J.~M. Lasry and P.~L. Lions, ``Jeux {\`{a}} champ moyen. i - le cas
  stationnaire,'' \emph{Comptes Rendus Math\'ematique}, vol. 343, no.~9, pp.
  619--625, 2006.

\bibitem{LAS06b}
------, ``Jeux {\`a} champ moyen. ii - horizon fini et contr{\^o}le optimal,''
  \emph{Comptes Rendus Math\'ematique}, vol. 343, no.~10, pp. 679--684, 2006.

\bibitem{GUE10}
O.~Gu{\'e}ant, J.~M. Lasry, and P.~L. Lions, ``Mean field games and oil
  production,'' \emph{Preprint}, 2010.

\bibitem{GUE11}
------, ``Mean field games and applications,'' \emph{Paris-Princeton Lectures
  on Mathematical Finance 2010}, pp. 205--266, 2011.

\bibitem{kotelenez2010macroscopic}
P.~Kotelenez and T.~Kurtz, ``Macroscopic limits for stochastic partial
  differential equations of mckean--vlasov type,'' \emph{Probability theory and
  related fields}, vol. 146, no.~1, pp. 189--222, 2010.

\end{thebibliography}

\end{document}